\documentstyle[prc,aps,twocolumn]{revtex}

\begin{document}
\draft

\input{psfig}

\twocolumn[
\hsize\textwidth\columnwidth\hsize\csname @twocolumnfalse\endcsname

\title{Radiative capture of protons by deuterons}

\author{W.~Schadow}
\address{
Institute of Nuclear and Particle Physics,  and
Department of Physics,  Ohio University, Athens, OH 45701}

\author{W.~Sandhas}
\address{Physikalisches	 Institut der Universit\"at Bonn, Endenicher
Allee 11-13, D-53115 Bonn, Germany}

\vspace{10mm}

\date{August 17, 1998}

\maketitle

\begin{abstract}
The differential cross section for radiative capture of protons by
deuterons is calculated using different realistic NN interactions. We
compare our results with the available experimental data below
$E_x$~=~20~MeV. Excellent agreement is found when taking into account
meson exchange currents, dipole and quadrupole contributions, and the
full initial state interaction. There is only a small difference
between the magnitudes of the cross sections for the different
potentials considered. The angular distributions, however, are
practically potential independent.
\end{abstract}


]


The radiative capture of protons by deuterons and the inverse
reaction, the photodisintegration of $^3$He, have been investigated
experimentally and theoretically over the last decades with quite some
interest. Despite the various corresponding investigations, the theory
is only in rough agreement with experiment, and there are
inconsistencies between the data up to 30\% in the magnitudes of the
cross sections. The experimental results by Belt et al. \cite{Belt70}
and King et al. \cite{King84a,King84b} are in good agreement. Those by
Matthews et al. \cite{Matthews74} and Skopik et al. \cite{Skop83a}
agree in the angular distributions, but disagree in the magnitudes of
the cross sections. This indicates a calibration problem of the
measurements.

From the theoretical side several attempts have been made to describe
the cross sections in this energy region. In the early calculations by
Barbour et al. \cite{Barb70} phenomenological interactions were used.
It was shown that the final state interaction is quite important, and
that the E2 contributions in the electromagnetic interaction are
needed in the differential cross section. In the calculations by
Gibson and Lehman \cite{Gibs75} a more realistic Yamaguchi
interaction, but only the E1 components were employed.
King~et~al.~\cite{King84a} performed an effective two-body, direct
capture calculation with the initial state being treated as a plane
wave, or as a scattering state generated from an optical
potential. Fonseca and Lehman \cite{Fons95a} calculated the
polarization observables $A_{yy}$ and $T_{20}$ at the excitation
energy $E_x$ = 14.75 MeV including only the E1 interaction. A
calculation at $E_x = 15$ MeV based on realistic interactions and
both, the E1 and E2 contributions has been done by Ishikawa and
Sasakawa \cite{Ishi92b}. Another calculation of $A_{yy}$ in this
energy region is by Jourdan et al. \cite{Jourd86}. It was found in all
these investigations that $T_{20}$ is independent of the deuteron and
the helium $D$-state probability, whereas $A_{yy}$ shows a weak
dependence on these quantities.

Very-low-energy radiative capture processes are of considerable
astrophysical relevance. The $p$-$d$ radiative capture, which at such
energies is almost entirely a magnetic dipole (M1) transition, was
studied in plane wave (Born) approximation by
Friar~et~al. \cite{Friar90}. In these investigations the authors
employed their configuration-space Faddeev calculations of the helium
wave function, with inclusion of three-body forces and pion exchange
currents.  Various trends, e.g., the correlation between cross sections
and helium binding energies, and their potential dependence were pointed
out.  More recently a rather detailed investigation of such processes
has been performed by Viviani et~al. \cite{Viviani96a}. Their
calculations employed the quite accurate three-nucleon bound- and
continuum states obtained in the variational pair-correlated
hyperspherical method, developed, tested and applied over years by
this group.

In Refs. \cite{Schadow98a,Sandhas98a} we have treated the $^3$He
photodisintegration and the inverse radiative capture process within
the integral equation approach discussed below.  These calculations
were based on the Paris, Bonn {\sl A}, and Bonn {\sl B} potentials in
Ernst-Shakin-Thaler (EST) representation: PEST, BAEST, BBEST
\cite{Haid84,Haid86a,Haid86b}. We have demonstrated in particular the
role of E2 contributions, meson exchange currents, and higher partial
waves at $E_x$ = 12 MeV and $E_x$ = 15 MeV. The sensitivity against
the underlying potentials, moreover, was pointed out. In the present
paper we extend these investigations and compare our calculations with
all sufficiently accurate data below $E_x$ = 20 MeV.

The Alt-Grassberger-Sandhas (AGS) equations are well known to go over
into effective two-body Lippmann-Schwinger equations \cite{Alt67} when
representing the input two-body {\em T}-operators in separable
form. The proton-deuteron scattering amplitude, thus, is determined
by

\begin{equation}
\label{scampl}
 {\cal T} ({\bf q}, {\bf q}'') = {\cal V} ({\bf q}, {\bf q} '')
 + \int \! d^{\,3} q' \: {\cal V} ({\bf q}, {\bf q}')\:{\cal G}_0 ({\bf q}')
\: {\cal T} ({\bf q}',{\bf q}'') .
\end{equation}

\noindent
Applying the same technique to the $^3$He photodisintegra-

\begin{figure}[hbt]
\begin{minipage}[t]{8.3cm}
\hspace{4mm}
\psfig{file=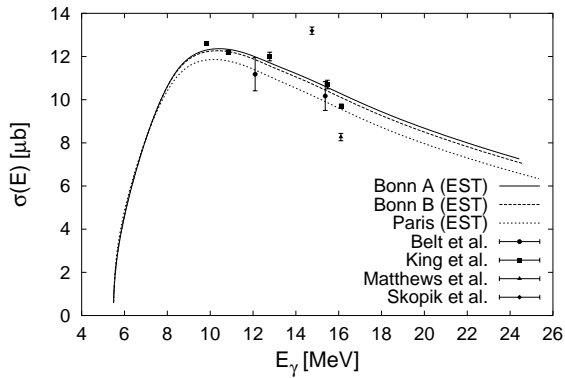,width=74mm,angle=-90}
\vspace{2mm}
\caption{Total cross section for the capture of protons by
deuterons. The  data are from [1-5].}
\end{minipage}
\vspace{4mm}
\end{figure}

\noindent
tion process, an integral equation of rather similar structure is
obtained for the corresponding amplitude \cite{Gibs75},

\begin{equation}
\label{pampl}
 {\cal M} ({\bf q} \,) = {\cal B} ({\bf q})
 + \int \! d^{\,3} q' \: {\cal V} ({\bf q}, {\bf q}')\:{\cal G}_0 ({\bf q}')
\: {\cal M} ({\bf q}') .
\end{equation}

In both equations the kernel is given by an effective proton-deuteron
potential ${\cal V}$ and an effective free Green function ${\cal
G}_0$.  However, in Eq. (\ref{pampl}) the inhomogeneity of
Eq. (\ref{scampl}) is replaced by an off-shell extension of the
$^3$He photodisintegration amplitude in plane-wave (Born) approximation,

\begin{equation}
\label{bampl}
 B ({\bf q}\,) = \: \langle{\bf q}\,|  \langle \psi_{d}|
H_{\mbox{\scriptsize em}}|\psi_{\rm He}
\rangle .
\end{equation}

\noindent
Here, $|\psi_{\rm He} \rangle$ and $| {\psi_d} \rangle$ are the $^3$He
and deuteron states, $|{{\bf q}}\rangle $ is the relative momentum
state of the proton, $H_{\mbox{\scriptsize em}}$ denotes the
electromagnetic operator. In other words, with this replacement any
working program for $p$-$d$ scattering, based on separable
representations or expansions of the two-body potential, can
immediately be applied to calculating the full $^3$He
photodisintegration amplitude with inclusion of the final-state
interaction.

The cross section for the $p$-$d$ capture process is obtained from the
corresponding photodisintegration expression by using the principle of
detailed balance \cite{Crave76}

\begin{equation}
\frac{d \sigma^{\rm dis}}{d \Omega} =
\frac{3}{2} \, \frac{k^2}{Q^2} \,
\frac{d \sigma^{\rm cap}}{d \Omega}.
\end{equation}

\noindent
Here, $k$ and $Q$ are the momenta of the proton and the photon,
respectively. In the present treatment no Coulomb forces have been
taken into account. The matrix element (\ref{bampl}) for $p$-$d$
capture differs from the corresponding $n$-$d$ expression only in its
isospin content.

The results presented in this paper are obtained by employing the
PEST, BAEST and BBEST potentials as input \cite{Haid84,Haid86a},
however, with an improved parameterization by Haidenbauer
\cite{Haidprivate}.  The high quality of this input has been
demonstrated in bound-state and scattering calculations
\cite{Haid86b,Corne90a,Park91}.

\begin{figure}[hbt]
\begin{minipage}[t]{8cm}
\hspace{4mm}
\psfig{file=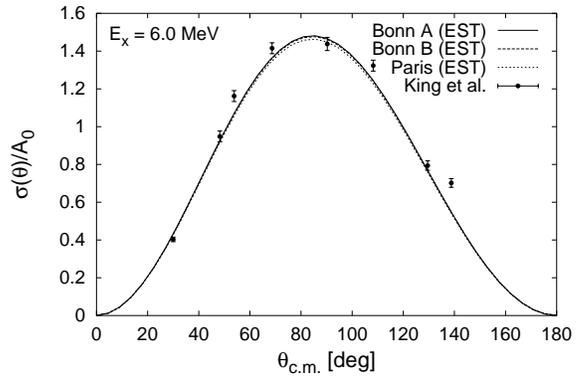,width=74mm,angle=-90}
\vspace{2mm}
\caption{Angular distribution of the cross section for $p$-$d$ capture
at $E_x$~=~6.0~MeV. The data are from [2,3].}
\end{minipage}
\end{figure}

 The electromagnetic operator relevant in the total cross section is,
at the low energies considered, essentially a dipole operator. In the
differential cross section we have to include also the quadrupole
operator.  According to Siegert's theorem \cite{Sieg37}, these
operators are given by

\begin{equation}
{ {H_{em}^{(1)}}}  \sim  - i \, E_\gamma\,
\sum_{i = 1}^3 e_i  \, r_i \, Y_{1 \lambda}(\vartheta_i, \varphi_i)
\end{equation}

\noindent
and

\begin{equation}
{ {H_{em}^{(2)}}} \sim \frac{E_\gamma^2}{\sqrt{20}}\,
\sum_{i = 1}^3 e_i  \, r_i^2 \, Y_{2 \lambda}(\vartheta_i, \varphi_i) ,
\end{equation}

\noindent
where $E_{\gamma}$ denotes the photon energy, $r_i$ the nucleon
coordinates, $e_i$ the electric charges, and $\lambda = \pm 1$ the
polarization of the photon.

Our method for determining the final state, i.e. the $^3$He wave
function, is described in
Refs. \cite{Cant97a,SchadowHaidsubmitted}. In the calculation of the
Faddeev components the total angular momentum $j$ of the two-body
potential was restricted to $j \leq 2$, while in the full state all
partial waves with $j \leq 4$ (34 channels) have been taken into
account.  With this number of channels a converged calculation was
achieved, incorporating 99.8\% of the wave functions. 

\begin{figure}[hbt]
\begin{minipage}[t]{8.3cm}
\hspace{2mm}
\psfig{file=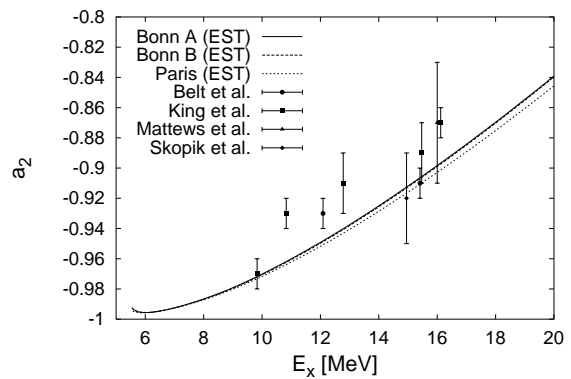,width=74mm,angle=-90}
\vspace{2mm}
\caption{The $a_2$ angular distribution coefficient as function of $E_x$. 
The data are from [1-5].}
\end{minipage}
\end{figure}

\twocolumn[\hsize\textwidth\columnwidth\hsize\csname
@twocolumnfalse\endcsname

{\begin{figure}[hbt]
\begin{minipage}[t]{17.9cm}
\hspace{6mm}
\psfig{file=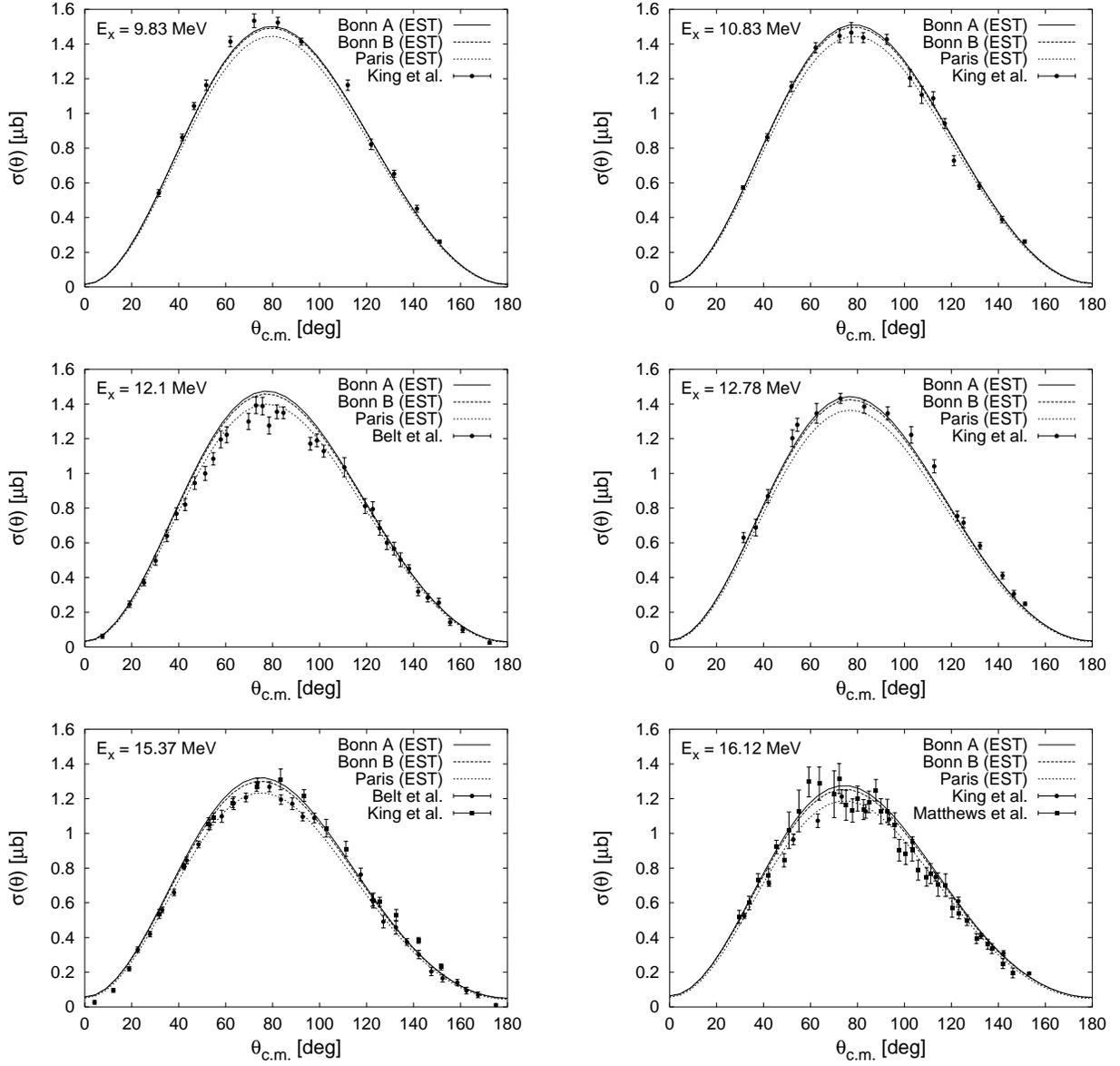,width=130mm}
\caption{Differential cross section for $p$-$d$ capture for energies
$E_x$ from near threshold up to 16 MeV. The  data are from [1-4].
The data set by Matthews et al. [4] has been renormalized with the $A_0$
from King et al. [2,3]. }
\end{minipage}
\end{figure}}
]

\noindent
Details concerning their high quality are given in
\cite{SchadowHaidsubmitted}. For the initial state all partial waves
with $j \leq 2$ have been included in order to get a converged
calculation of the cross section \cite{Schadow98a,Sandhas98a}.

Usually the differential cross section is expanded in terms of Legendre
polynomials

\begin{equation}
\label{eqlegendre}
\sigma (\theta) = A_0 \left ( 1 + \sum_{k = 1}^{4} a_k \, P_k (\cos \theta)
\right ).
\end{equation}

\noindent
The total cross section is obtained by integrating over the angle
$\theta$ between the incoming photon and the outgoing proton

\begin{equation}
\label{eqtotal}
\sigma  = 4 \pi \, A_0.
\end{equation}

Figure 1 shows the total cross sections for the Paris, the Bonn {\sl
A} and Bonn~{\sl B} potentials compared to the experimental data
\cite{Belt70,King84a,King84b,Matthews74,Skop83a}. There is a small
potential dependence of $\sigma$ and, hence, of $A_0$ similar to the
one observed in the corresponding photoprocess
\cite{Schadow98a,Sandhas98a}. In view of the errorbars, the
experimental data by Belt et al.  \cite{Belt70} and King et
al. \cite{King84a,King84b} are reproduced for all potentials with the
same quality. Those by Matthews et al. \cite{Matthews74} and by Skopik
et~al. \cite{Skop83a} are not described by the theoretical curves.

Figure 2 shows the angular distribution of the differential cross
section, i.e., the ratio of $\sigma(\theta)$ and the coefficient $A_0$
compared to the experimental data~\cite{King84b}. This distribution is
evidently potential independent. In other words, its shape shows no
correlation with the helium binding energy, or the $D$-state
probability of the $^3$He wave function.

Figure 3 shows the angular distribution coefficient $a_2$ of the
expansion (\ref{eqlegendre}) compared to the coefficients extracted
from experiment \cite{Belt70,King84a,King84b,Matthews74,Skop83a}. In
accordance with Figure 2 there is almost no potential dependence,
i.e. no dependence on the three-body binding energy and the $D$-state
probability, although this probability varies for the three potentials
considered between 6 to 8\% \cite{SchadowHaidsubmitted}.

Figure 4 shows the differential cross sections obtained for these
potentials at various energies compared to the experimental data. Due
to the slight potential dependence of the total cross section and,
thus, of $A_0$, the magnitudes of the curves differ correspondingly.
In all cases there is good agreement between theory and experiment.
As pointed out in \cite{Schadow98a,Sandhas98a} this agreement can only
be achieved by taking onto account E1 and E2 contributions of the
electromagnetic interaction, meson exchange currents, and higher
partial waves in the potential and in the three-body wave function.
It should be mentioned that for increasing energies the peak is
slightly shifted to the right-hand side, because of a smaller E1 and a
somewhat higher E2 contribution. Note that, due to the missing E1-E2
interference term, the quadrupole contribution is practically
negligible in the total cross section.

In \cite{Schadow98a,Sandhas98a} we have shown that for different
potentials the low-energy peak heights of the $^3$He
photodisintegration cross sections are strictly correlated with the
corresponding $^3$He binding energies, and with the number of partial
waves included. The magnitude of the present radiative capture
process, i.e., the constant $A_0$, appears to be similarly fixed by
the three-body binding energy. In other words, at the energies
discussed, the radiative capture cross section does not represent an
additional observable for testing different potentials.

\acknowledgements We acknowledge financial support from the Deutsche
Forschungsgemeinschaft under Grant No. Sa 327/23-I.  Part of this work
was performed in part under the auspices of the U.~S.  Department of
Energy under contract No. DE-FG02-93ER40756 with Ohio University.  The
authors would like to thank H.~R.~Weller for providing additional
information about the experimental data.

\end{document}